\documentclass[12pt,preprint]{aastex}

\usepackage[]{aastexug,natbib,epsfig,rotating,graphicx}

\def\bm{\bigskip}
\def\micron{$\mu$m}
\def\ptsec{$''\mskip-7.6mu.\,$}
\def\ptmin{$'\mskip-5.0mu.\,$}

\def\etal{et al.}
\def\ltsim{\raisebox{-.4ex}{$\stackrel{<}{\sim}$}}
\def\gtsim{\raisebox{-.4ex}{$\stackrel{>}{\sim}$}}
\def\arcsec{$^{\prime\prime}$}
\def\arcmin{$^{\prime}$}
\def\twocolheads#1{\multicolumn{2}{c}{#1}}

\def\et{\bm\end{table}}
\def\bt{\begin{table}[htbp]\bm}
\def\efig{\end{figure}}
\def\bfig{\begin{figure}[htb]\bm}

\slugcomment{Accepted by ApJ:  September 12, 2005}

\shorttitle{Deep Near-Infrared Observations of L\,1014}
\shortauthors{Huard et al.}

\begin{document}


\title{Deep Near-Infrared Observations of L\,1014: Revealing the nature of the core and its embedded source}


\author{Tracy L. Huard, Philip C. Myers}
\affil{Harvard-Smithsonian Center for Astrophysics, 60 Garden Street,
	Cambridge, MA 02138}
\email{thuard@cfa.harvard.edu, pmyers@cfa.harvard.edu}

\author{David C. Murphy}
\affil{Carnegie Institution of Washington, 813 Santa Barbara Street,
	Pasadena, CA 91101}
\email{david@ociw.edu}

\author{Lionel J. Crews}
\affil{Department of Geology, Geography, and Physics,
	University of Tennessee at Martin, Martin, TN 38238}
\email{lcrews@utm.edu}

\author{Charles J. Lada, Tyler L. Bourke}
\affil{Harvard-Smithsonian Center for Astrophysics, 60 Garden Street,
	Cambridge, MA 02138}
\email{clada@cfa.harvard.edu, tbourke@cfa.harvard.edu}

\author{Antonio Crapsi}
\affil{Osservatorio Astrofisico di Arcetri, Largo Enrico Fermi 5, I-50125 Firenze, Italy}
\email{crapsi@arcetri.astro.it}

\author{Neal J. Evans II}
\affil{Department of Astronomy, University of Texas at Austin,
	1 University Station C1400, Austin, TX 78712-0259}
\email{nje@astro.as.utexas.edu}

\and

\author{Donald W. McCarthy, Jr., Craig Kulesa}
\affil{Steward Observatory, University of Arizona, 933 North Cherry Avenue
	Tucson, AZ 85721}
\email{mccarthy@as.arizona.edu, ckulesa@as.arizona.edu}




\begin{abstract}

Recently, the Spitzer Space Telescope discovered L\,1014-IRS, a mid-infrared source
with protostellar colors, toward the heretofore ``starless'' core L\,1014. 
We present deep near-infrared observations that show a scattered light nebula
extending from L\,1014-IRS.  This nebula resembles those typically 
associated with protostars and young stellar objects, tracing envelope cavities
presumably evacuated by an outflow.  The northern lobe of the nebula has an opening
angle of $\sim$100$^\circ$, while the southern lobe is barely detected.  Its morphology
suggests that the bipolar cavity and inferred protostellar disk is not inclined more than
30$^\circ$ from an edge-on orientation.  The nebula extends at least 8\arcsec\ from the
source at K$_s$, strongly suggesting that L\,1014-IRS is embedded within L\,1014 at a distance
of 200 pc rather than in a more distant cloud associated with the Perseus arm at 2.6 kpc.
In this case, the apparently low luminosity of L\,1014-IRS, 0.090~L$_\odot$, is consistent with
it having a substellar mass.  However, if L\,1014-IRS is obscured by a circumstellar disk,
its luminosity and inferred mass may be greater.
Using near-infrared colors of background stars, we investigate
characteristics of the L\,1014 molecular cloud core.  We determine a
mass of 3.6~M$_\odot$ for regions of the core
with $A_V \ge$ 2 magnitudes.  A comparison of the radial extinction profile of L\,1014 with
other cores suggests that L\,1014 may be among the most centrally condensed cores known,
perhaps indicative of the earliest stages of brown dwarf or star formation processes.

\end{abstract}




\section{INTRODUCTION}\label{intro}

The discovery by the Cores-to-Disks (hereafter, {\it{c2d}}; Evans et al. 2003) Spitzer
Legacy team of L\,1014-IRS (SSTc2d 2124075$+$495909; Young et al. 2004), a
mid-infrared source with protostellar colors (Allen et al. 2004; Megeath et al. 2004)
observed toward the dark cloud L\,1014, was unexpected since L\,1014 was previously
classified as a starless core based on lack of an associated IRAS source (Lee \& Myers 1999).
Modeling the mid-infrared through mm-wave spectral energy distribution (SED), Young
et al. (2004) found that L\,1014-IRS was best fitted with a total protostar and disk
luminosity of 0.09~L$_\odot$, assuming it was embedded within L\,1014 at 200 pc.
In that case, L\,1014-IRS would be either the lowest luminosity protostar or embedded
brown dwarf known.  However, the nature of the source was ambiguous since a more
distant cloud associated with the Perseus arm at 2.6 kpc exists along the same line of
sight as L\,1014.  If the protostar were instead embedded in the distant cloud, then its
luminosity would be similar to that of a typical T Tauri star.  Determination of the distance
to L\,1014-IRS is important since, if it is indeed associated with the nearby L\,1014
core, then it apparently has a remarkably low mass.

The most unambiguous determination of its distance would be to detect a
protostellar outflow, either at the radial velocity associated with the
nearby L\,1014 core or the more distant cloud.
In an attempt to detect an outflow from L\,1014, Crapsi et al. (2005b)
mapped the core in CS(2-1) and CO(1-0) with the Five College
Radio Astronomy Observatory (FCRAO).
The CS(2-1) spectra do not indicate an asymmetric
double-peaked feature, with the blue peak stronger than the red peak, typically
used to identify infall motions arising from the star formation process in cores
(e.g., Lee, Myers, \& Tafalla 1999).  Furthermore, the study found no significant
wing-like CO emission arising from a molecular outflow.
Crapsi et al. (2005b) concluded that their data do not
indicate any classical outflow signatures, at least for angular scales 
of $\gtsim$45\arcsec, the size of the FCRAO beam.  If L\,1014-IRS is associated
with the nearby core, and therefore of very low luminosity, sensitive submillimeter
interferometric observations probing much smaller angular scales
may be necessary to detect the outflows from such sources.

In this paper, we present deep near-infrared observations showing nebulosity
around L\,1014-IRS strongly suggesting that L\,1014-IRS is indeed associated
with L\,1014.  We describe our observations in \S\ref{obs}.  In \S\ref{proto}, we
discuss the properties of L\,1014-IRS as indicated by these observations.
Finally, in \S\ref{core}, we construct an extinction map of the core, derive its
mass, and discuss the core density profile.

\section{OBSERVATIONS}\label{obs}

As part of a larger program to survey the {\it{c2d}} cores in the near-infrared, we
obtained deep J, H, and K$_s$ observations of L\,1014 during UT 29--31 July 2004 using
PISCES (McCarthy et al. 2001) on the converted Multiple Mirror Telescope (MMT) and
during UT 24 September 2004
using FLAMINGOS (Elston 1998) on the 4-meter telescope at Kitt Peak National Observatory (KPNO).
With a $\sim$10\arcmin$\times$10\arcmin\ field of view, the purpose of the FLAMINGOS
observations was to obtain deep observations of the entire L\,1014 core.  The PISCES
observations, with a circular field of view of $\sim$3\ptmin 1 on the MMT, were obtained to
provide even deeper observations of L\,1014-IRS
and the center of the core.  Seeing in the FLAMINGOS images was $\sim$0\ptsec 9.
In the PISCES images, the seeing varied from 0\ptsec 4--1\ptsec 0, depending on the night,
but typically was $\sim$0\ptsec 7.

The L\,1014 core was observed with FLAMINGOS for a total of 6 minutes at J,
168 minutes at H, and 33 minutes at K$_s$.  These times were obtained by combining
dithered images with individual exposure times of 60 seconds at J and H,
and 15 seconds at K$_s$.  Since these exposure times saturated moderately
bright stars, we obtained suites of at least 9 dithered, 5-second images 
from which the photometry of most of these sources
could be derived.  Each individual image was reduced using IRAF and 
IDL procedures following a standard method for reducing near-infrared images.
Specifically, (1) a dark image, representing the signal due to the
detector bias and temperature, was subtracted from the raw image, (2) the
dark-subtracted image was then divided by a dome flat field to account
for the nonuniform pixel-to-pixel sensitivity across the detector,
and (3) the sky was subtracted.  The individual images were then
aligned and averaged, with discrepant pixel values rejected, to produce
the final image at J, H, and K$_s$.

Sources in these final FLAMINGOS images were detected and aperture
photometry obtained using {\it{PhotVis}} Version 1.09 (Gutermuth et al. 2004),
an IDL GUI-based aperture
photometry tool, which makes use of the standard
{\it{daophot}} packages (Landsman 1993).
The positions of these sources were derived by comparison with known
positions of stars, as listed in the USNO-A2.0 catalog, that were observed in
our fields.  Instrumental magnitudes at J, H and K$_s$, obtained from
1\ptsec 0-radius synthetic apertures centered on these sources, were
calibrated using the published J, H, and K$_s$ magnitudes of stars within
our fields that were listed in the 2MASS catalog.

Our final L\,1014 catalog contains 7000 sources detected at
both H and K$_s$, with most of these sources having H and K$_s$ photometric uncertainties
less than 0.15~mag.  We estimate our 90\% completeness limits
to be $\sim$18.5, $\sim$19.0 and $\sim$18.0 at J, H, and K$_s$, repectively.
Uncertainties in the positions of individual sources are less than 1\arcsec.

The central region of L\,1014 was observed with PISCES for a total of
13.5 minutes at J, 68 minutes at H, and 9 minutes at K$_s$.  These times were
obtained by combining randomly dithered images with individual exposure times
of 30 seconds at J and H, and 10 seconds at K$_s$.  These images
were supplemented with at least 9 dithered, 0.75-second
images and 9 dithered, 5-second images at each band from which the photometry of
the bright, saturated sources could be derived.

The individual PISCES images were corrected for quadrant cross-talk effects present
in HAWAII arrays (McCarthy et al. 2001).  Then, each image was reduced
following the standard method previously
described, except twilight flat fields were used instead of dome flats.
In addition, PISCES images were corrected for geometric distortion prior to 
being aligned and averaged together.

\section{THE L\,1014-IRS PROTOSTAR AND NEBULA}\label{proto}

The position of L\,1014-IRS, as reported by Young et al. (2004), is
$\alpha = $21$^h$24$^m$07.51$^s$, $\delta = +$49$^\circ$59\arcmin 09\ptsec 0
(J2000.0).  This position matches the coordinates of a very red source detected
at K$_s$, seen near the center of the K$_s$ image in Figure~\ref{L1014-IRS_HK},
to within 0\ptsec 5.  This separation is within our astrometric tolerance; thus,
this K$_s$ source is consistent with being the near-infrared counterpart to
L\,1014-IRS.

\subsection{Detection and Inclination of the Cavity}\label{inclination}

\begin{figure*}[tbh]
\epsscale{1}   
\plotone{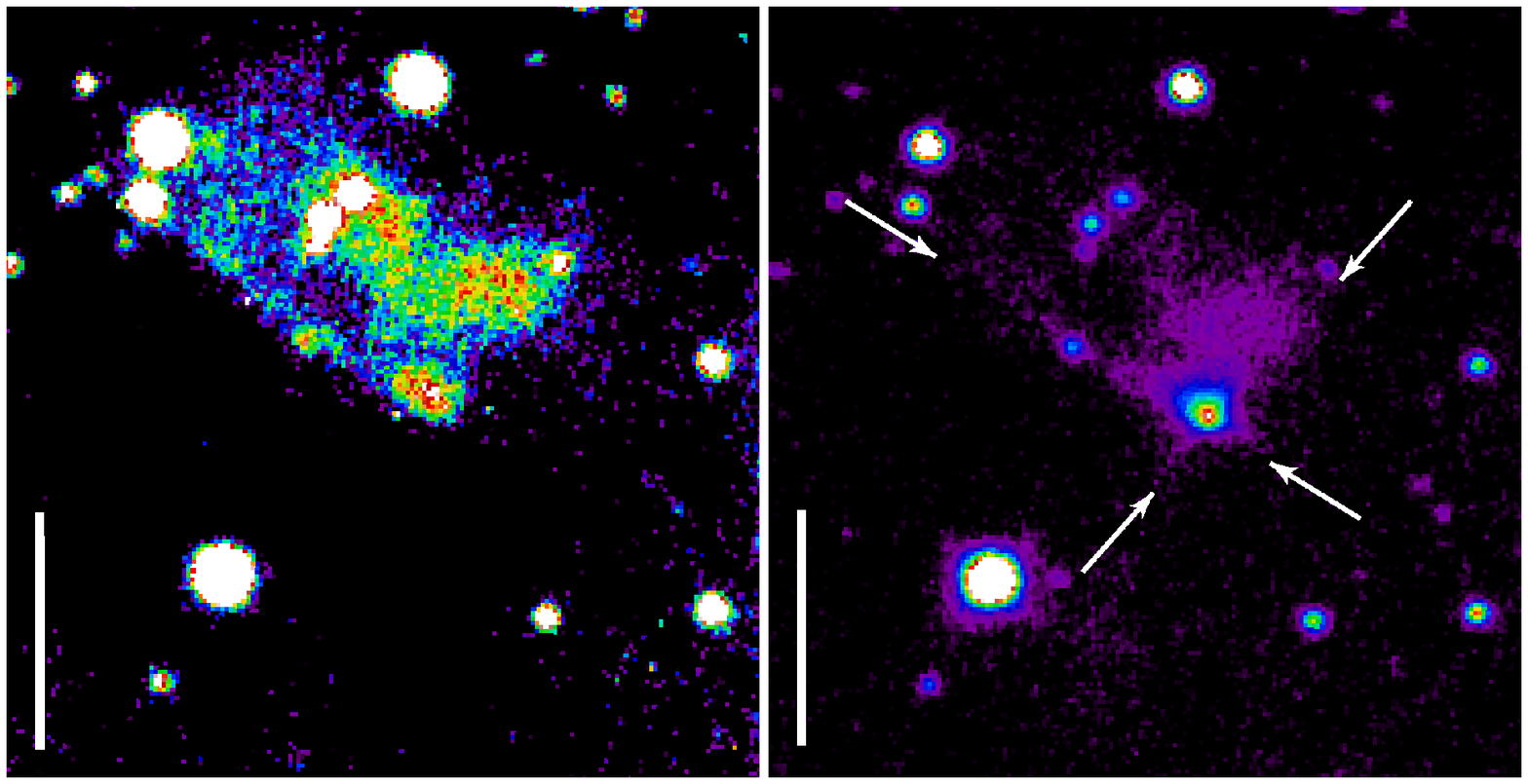}
\caption{Deep H (left) and K$_s$ (right) images of L1014-IRS obtained with PISCES on the MMT,
in which the intensity is scaled to enhance the nebulosity from the
cavity north of L\,1014.  The arrows overlaid on the K$_s$ image point along the projected
boundaries of the conical cavity.  Faint nebulosity can be seen south of the arrow identifying
the southeastern boundary, suggesting that the southern lobe may exhibit some curvature.
The white bar in the bottom left corner of each image represents
an angular size of 10\arcsec\ (0.01 pc).
\label{L1014-IRS_HK}}
\end{figure*}

\begin{figure*}[tbh]
\epsscale{1}  
\plotone{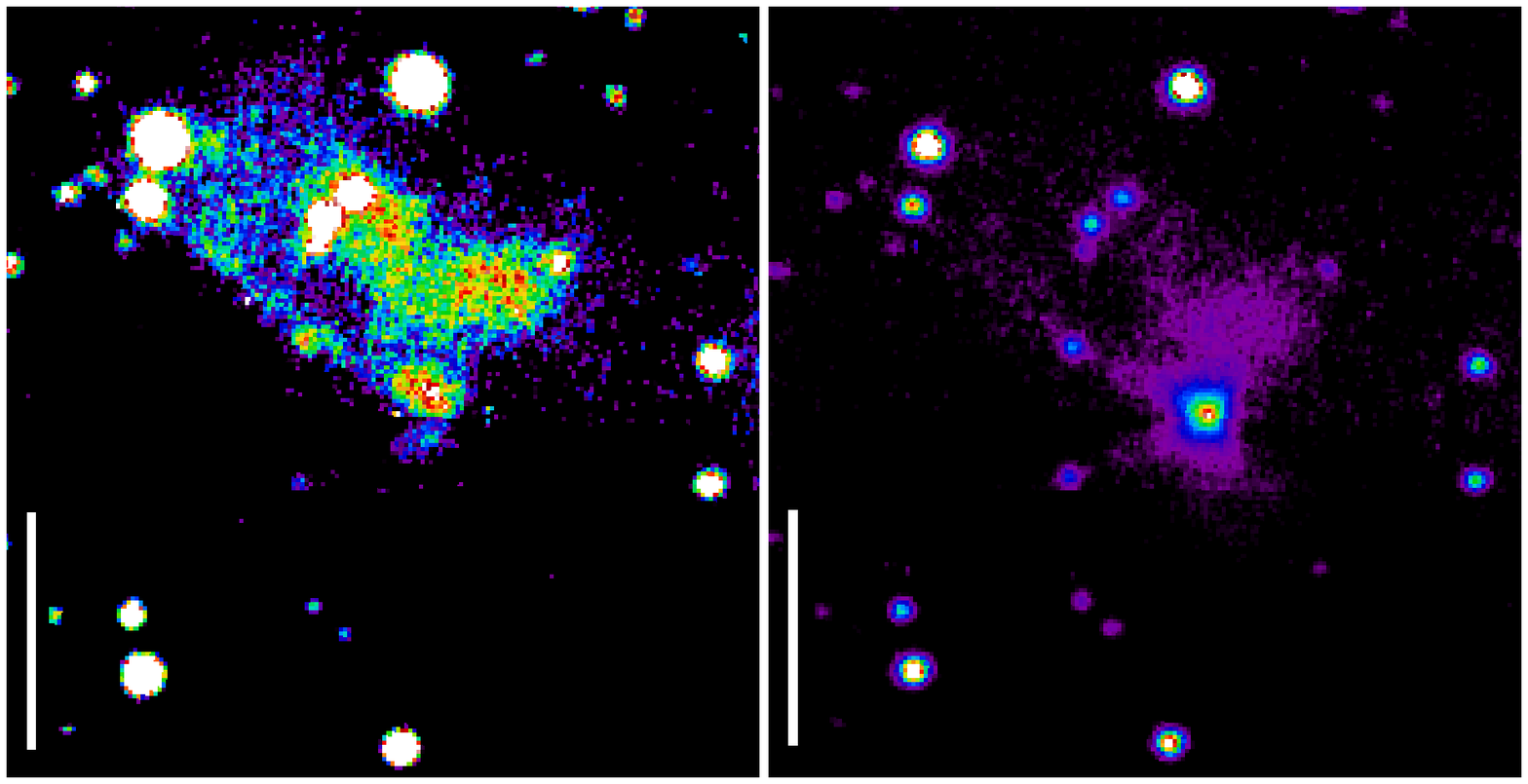}
\caption{H (left) and K$_s$ (right) images of L1014-IRS expected from an edge-on 
bipolar cavity surrounding L\,1014-IRS, accounting for differences in foreground extinction
from the L\,1014 core toward the northern and southern lobes.  These images were constructed
by reflecting pixels north of L\,1014-IRS about the horizontal line through L\,1014-IRS and
then applying additional extinction to the southern pixels (see \S\ref{inclination}).
Hence, stars south of L\,1014-IRS appearing in these images are artifacts of 
the construction.  The white bar in the bottom left corner of each image represents
an angular size of 10\arcsec\ (0.01 pc).
\label{L1014-IRS_HKreflect}}
\end{figure*}

We detect a conical nebula, with opening angle of $\sim$100$^\circ$, extending
at least 8\arcsec\ north of L\,1014-IRS
at K$_s$, shown in Figure~\ref{L1014-IRS_HK}.
The north and northwestern portions of this nebula appears brighter at K$_s$ than the
northeastern portion.  A more extended nebula is observed at H, also resembling a conical
or fan-shaped nebula with L\,1014-IRS at the apex.  Similar to the K$_s$ nebulosity, the 
nebulosity at H is brighter in the north and northwestern portions, but peaks somewhat
farther from L\,1014-IRS than the does the K$_s$ nebulosity.  Such features are typical
of scattered light nebulae associated with bipolar cavities emanating from
young stellar objects (YSOs) and protostars
(e.g., Duch\^ene et al. 2004; Park \& Kenyon 2002;
Padgett et al. 1999; Whitney, Kenyon, \& Go\'mez 1997).  Although the brightest 
H nebulosity is seen in the north and northwest portions of the cavity, the nebula
is most extended in the northeastern half of the cavity, at least 15\arcsec\ from
L\,1014-IRS.

The nebulosity near L\,1014-IRS
is dominated by the northern lobe, but there is a hint of nebulosity at K$_s$ associated with
the projected boundaries of the southern lobe within $\sim 2-3$\arcsec\ south
of L\,1014-IRS.  The projected boundaries of the
southern lobe appear to be nearly extensions of those of the northern lobes, as illustrated in
the figure, though there seems to be some hour-glass-like curvature associated with
the nebula if faint nebulosity located 8--10\arcsec\ southeast of L\,1014-IRS is tracing
that projected boundary.

The prominence of the nebulosity associated with the northern lobe compared with
that of the southern lobe is at least partly explained by extinction effects.
Assuming that L\,1014-IRS is embedded in the nearby L\,1014 core (see \S\ref{distance}),
with half of the core material 
in the foreground and half background to L\,1014-IRS, the steeply increasing extinction
from the northern to the southern regions
of the L\,1014 core (see \S\ref{map}) suggests that light scattered from the southern
lobe suffers an {\it{additional}} extinction of $A_V \approx$ 5-10 magnitudes compared
to light scattered from the corresponding position in the northern lobe.  With a standard
reddening law (e.g., Rieke \& Lebofsky 1985), if the plane of symmetry of L\,1014-IRS 
is oriented edge-on, 
then H light scattered from the southern lobe would be only 20--45\% as bright as the northern
lobe.  At K$_s$, the southern lobe would be only 35--60\% as bright as the northern lobe.
At both H and K$_s$, most of the scattered light from the southern lobe would not be detected
in our images, except within $\sim$2--5\arcsec\ south of L\,1014-IRS.  

We explicitly demonstrate
this point by constructing H and K$_s$ images we would expect from an edge-on bipolar
cavity, accounting for this asymmetric extinction from the L\,1014 core.  To do so, we first construct
H and K$_s$ images where the pixels south of L\,1014-IRS were replaced with pixels
representing the reflection of the northern pixels about the horizontal line through
L\,1014-IRS.  Then, making use of the extinction map generated in \S\ref{map}, each southern
pixel value was diminished by an appropriate amount, depending on the additional extinction
at that southern position compared to the extinction at the corresponding northern
position.  Finally, noise was added to the southern pixel in order to ensure noise
characteristics in the southern pixels were similar to the northern pixels.
The resulting synthetic images expected from L\,1014-IRS, if oriented with an
edge-on inclination and accounting for asymmetric extinction from the core, are shown in
Figure~\ref{L1014-IRS_HKreflect}.  Comparing Figure~\ref{L1014-IRS_HKreflect} with
Figure~\ref{L1014-IRS_HK}, we see that the extinction from the core can explain
most of the lack of nebulosity from the southern lobe, but the observations do not
detect as much nebulosity as we might expect from an edge-on inclination, in particular
within $\sim$2--5\arcsec.

After accounting for extinction, an inclination of the bipolar cavity,
with the northern lobe pointing toward the observer and the southern lobe pointing
away, most naturally explains the relatively faint 
nebulosity observed toward the southern lobe.  In this way, only back-scattered light
will be detected from the southern lobe and, 
since dust in protostellar envelopes is believed to be more forward-scattering than
back-scattering (e.g., Draine \& Lee 1984; Kim, Martin, \& Hendry 1994;
Stark et al. 2005), nebulosity associated with this lobe would be fainter than
the northern nebulosity.
Radiative transfer models of protostellar environments (Stark et al. 2005) demonstrate
the differences in nebulosity expected for different inclinations, ranging from
an edge-on orientation ($i = 90^\circ$) to a pole-on orientation ($i = 0^\circ$).
These models suggest that inclinations in the range, $i \approx\ 60-90^\circ$,
are likely to show at least a hint of nebulosity from the lobe pointed away
from the observer.  Within this range, the less inclined cavities ($i \approx\ 60-75^\circ$)
exhibit only very faint nebulosity associated with the lobe pointed away, while
cavities with more edge-on inclinations ($i \approx\ 75-90^\circ$) exhibit
similarly bright nebulosity for both lobes.  Based on these models, the inclination
of L\,1014-IRS can be constrained conservatively to be within 30$^\circ$ of an
edge-on orientation.

\subsection{The Distance Ambiguity}\label{distance}

Whether the protostar is embedded within the nearby L\,1014 core or embedded within
the more distant cloud associated with the Perseus arm at 2.6~kpc (Brand \& Blitz 1993)
has been ambiguous, given the data available.  Such ambiguity in distance has 
a wide range of implications for the nature of L\,1014-IRS.

Assuming that L\,1014-IRS is embedded in the nearby L\,1014 core, located
at a distance of $\sim$200~pc, Young et al. (2004) find that the total luminosity of
the protostar and protostellar disk is $\sim$0.090~L$_\odot$.  They mention that the
partition of this luminosity between the protostar and disk is uncertain, 
but a protostar with photospheric luminosity 0.025~L$_\odot$ best fits the SED.
If we assume the protostar has a luminosity between 0.025--0.050~L$_\odot$,
evolutionary tracks of D'Antona \& Mazzitelli (1998, 1997; hereafter, DM97/98) imply a mass
of 30--45 M$_J$ for L\,1014-IRS assuming a typical Class I age of 10$^5$ years.  For
comparison, extrapolating the Lyon98/00 tracks\footnote{DM97/98 and Lyon98/00 tracks
are available online at
{\url{http://www.mporzio.astro.it/\~{}dantona/prems.html}} and
{\url{ftp://ftp.ens-lyon.fr/pub/users/CRAL/ibaraffe/BCAH98\_iso.1}}, respectively}
(Chabrier et al. 2000; Baraffe et al. 1998) to this 
age results in a similar substellar mass, 20--25 M$_J$.  If L\,1014-IRS is younger than
10$^5$ years, these mass estimates represent upper limits.  We caution, however,
that masses derived from current evolutionary tracks for young sources with ages less than 
10$^6$ years are rather uncertain (Baraffe et al. 2002).

Alternatively, if L\,1014-IRS is associated with the more
distant core, the luminosity of the protostar and protostellar disk would be $\sim$16~L$_\odot$
(Young et al. 2004).
In this case, referring to the evolutionary tracks of Siess, Dufour, \ Forestini (2000), the protostar would be
more akin to a $\sim$0.5--2~M$_\odot$ T Tauri star than a substellar object, depending
on the specific partitioning of flux between the protostar and disk.  Scattered light nebulae
detected around such YSOs of comparable or less luminosity (or mass) typically extend 
no more than 1500~AU from the source at K, including those with extended edge-on
morphologies (e.g., Duch\^ene et al. 2004; Padgett et al. 1999).  This estimate is derived from deep 
near-infrared observations of Taurus and Ophiuchus sources at 140~pc.  If these sources
were moved to 2.6~kpc and placed behind a core with peak extinction A$_V\gtsim 50$ (as 
observed in the nearby L\,1014 core; see \S\ref{map}), the detectable regions of these
nebulae likely would be smaller than 1500~AU.  Even so, the K$_s$ nebulosity detected in
our images extends at least 8\arcsec\ north of L\,1014-IRS, corresponding to at least 20000~AU
at 2.6~kpc.  In this scenario, this nebula would be at least an order of magnitude
larger than nebulae around similar YSOs.  Thus, {\it{the angular extent of the near-infrared
nebulosity detected in our observations strongly suggests that L\,1014-IRS is associated
with the nearby L\,1014 core, not the distant cloud}}.

\subsection{Significant Component of Scattered Light}

\begin{figure}[tbh]
\epsscale{0.45}   
\plotone{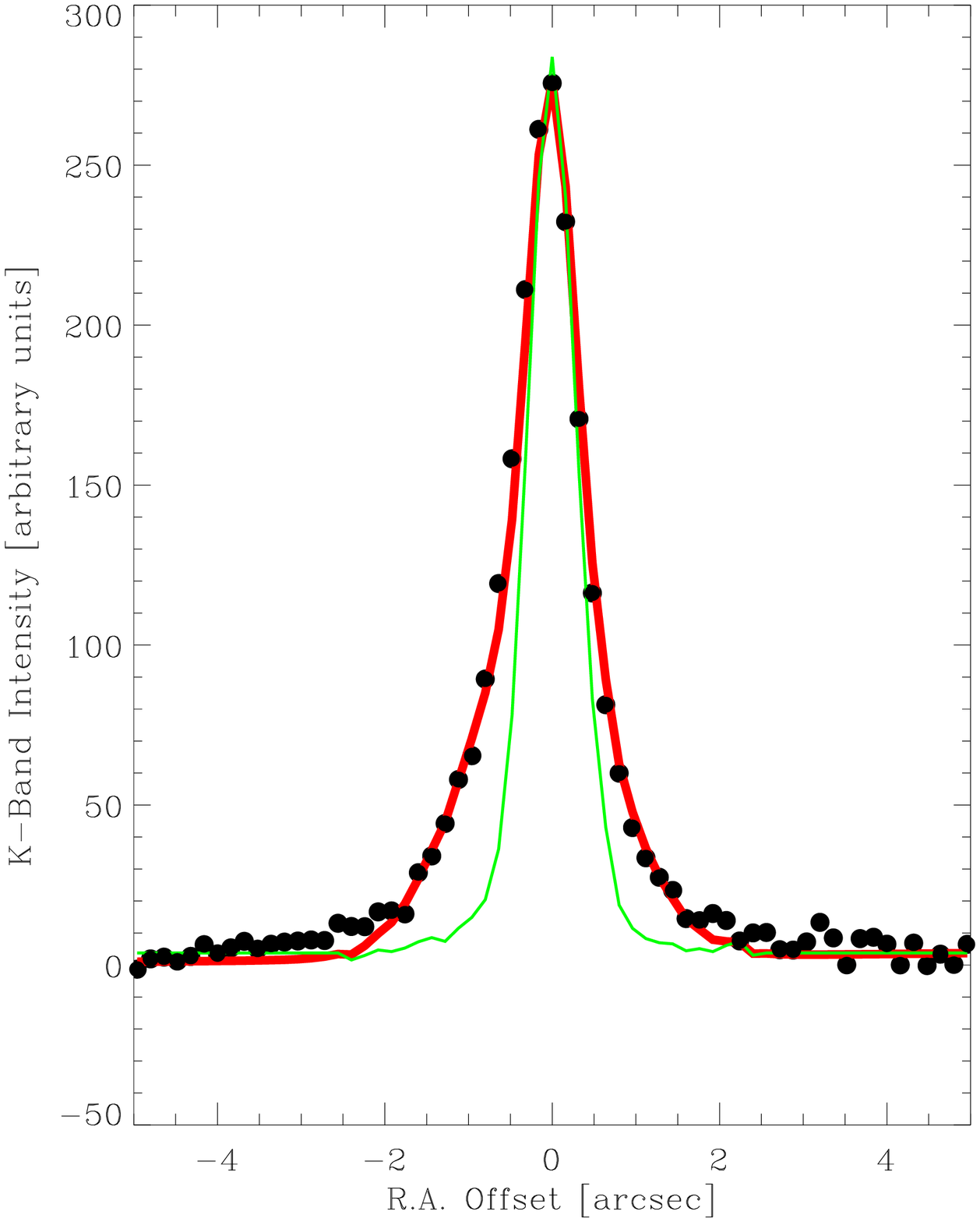}
\caption{K$_s$-band intensity cut through L\,1014-IRS along the east-west direction is plotted as black
data points.  This cut was derived by centering a slit of width 0\ptsec5 (3 pixels) and
length 20\arcsec\ on L\,1014-IRS.  The mean intensity of each column of three pixels was computed
and plotted as a data point.  The L\,1014-IRS profile is clearly broader than the PSF, plotted as a thin green
curve.  Assuming
the L\,1014-IRS profile is composed of  direct (extincted) light from the protostar (i.e., a PSF profile) and a broader Gaussian profile of scattered light, a fit to the observed profile is shown by the thick red curve.
\label{KcutEW}}
\end{figure}

Using a 1\arcsec-diameter aperture, we find that L\,1014-IRS has a K$_s$ flux of 0.20 $\pm$ 0.01~mJy,
or K$_s$ = 16.3 $\pm$ 0.1~mag, where calibration is the dominant source of uncertainty.  This detection
potentially has two components: direct, but extincted, light from the protostar and scattered protostellar
light.

We plot in Figure~\ref{KcutEW} an east-west cut of the K$_s$ intensity profile through
L\,1014-IRS.  This particular cut is the most narrow (or nearly so) K$_s$ profile for L\,1014-IRS,
and therefore will contain the least contamination due to scattered light.
Next, we constructed a typical PSF (normalized to unity) by taking the average of normalized PSFs for
five stars within 30\arcsec\ of L\,1014-IRS.  This typical PSF was then scaled to best-fit the K$_s$
flux distribution near the L\,1014-IRS peak.  The observed profile of L\,1014-IRS, however, is clearly
broader than the typical PSF, as shown in Figure~\ref{KcutEW}, suggesting that a significant amount
of scattering is observed along the line-of-sight toward L\,1014-IRS.

It is possible that the observed profile is {\it{entirely}} scattered light and, if so, then its deconvolved
size of 0\ptsec7 (140~AU) would correspond to the size of the primary scattering region.
In this case, our observations place no constraints on the K$_s$ protostellar flux.

If instead the profile is comprised of a typical PSF from the attenuated light from the protostar
and a broader Gaussian profile from the scattered light, then we find that {\it{at least}} 60\%
of the K$_s$ flux arises from scattering within our 1\arcsec-diameter aperture.  From this
consideration, if we have detected some direct light, the protostar would not be brighter
than K$_s \approx$ 17.3 magnitudes, and could be much fainter.  We do not detect L\,1014-IRS
in our deep H image; however, we derive a flux upper limit of H $>$ 18.5 magnitudes.  With our
upper limit on the K$_s$ protostellar flux of 17.3 magnitudes, this non-detection can be
explained by extinction from foreground dust in the core and extinction provided by a 
protostellar disk.

Regardless of whether we have detected entirely scattered light at K$_s$ or attenuated
protostellar light within brighter nebulosity, there is likely a significant contribution
of unresolved scattered light in the observed IRAC mid-infrared fluxes as well, even though 
those observations were not sensitive to the faint, extended nebulosity.
Based on the IRAC mid-infrared fluxes, Young et al. (2004)
estimated that L\,1014-IRS was best characterized with temperature of 700 $\pm$ 300~K
and photospheric luminosity of 0.025~L$_\odot$.  If, however, these mid-infrared fluxes were 
``contaminated'' by scattered light, in particular the 3.6~\micron\ and 4.5~\micron\
fluxes, then the temperature and photospheric luminosity may be somewhat greater.
For example, substellar luminosities were cited for HH~30 and Haro~6-5B until high
angular resolution, near-infrared Hubble Space Telescope (HST)
observations revealed that these sources are surrounded by nearly edge-on
disks, rendering only scattered circumstellar light visible at these wavelengths
(e.g., Padgett et al. 1999; Burrows et al. 1996).  Similarly, HL Tau had been thought
to be an optically visible classical T Tauri star until HST observations revealed that
it is also surrounded by a nearly edge-on disk and thus only seen in scattered light
at optical wavelengths (Stapelfeldt et al. 1995).  Thus, HH~30, Haro~6-5B, and HL~Tau
are characterized by greater photospheric luminosities than originally thought.

If L\,1014-IRS is obscured by a circumstellar disk in our observations, then
careful two-dimensional radiative transfer modeling of the L\,1014-IRS SED is
necessary to constrain its photospheric temperature and luminosity.
Taking into consideration scattered light, Whitney et al. (2004) developed two-dimensional
radiative transfer models to generate IRAC mid-infrared color-color and color-magnitude plots
expected for protostars of different temperatures and inclinations for the protostellar disk.
After accounting for extinction from the foreground portion of the L\,1014
core (i.e., $A_V \approx$ 25 magnitudes; see \S\ref{map}), the mid-infrared fluxes of L\,1014-IRS
indicate that this source has a temperature less than 4000~K, the lowest temperature
considered in those models, independent of the inclination.  Assuming an age of 10$^5$ years,
this temperature limit corresponds to a source with mass $\ltsim 0.4~M_\odot$ (DM97/98).
Depending on age and
mass, brown dwarfs may reach temperatures as high as $\sim$3000~K (e.g., Luhman et al. 2003;
DM97/98; Chabrier et al. 2000; Baraffe et al. 1998).  Clearly, a more detailed analysis of the
L\,1014-IRS SED with radiative transfer models that consider temperatures less than 4000~K
is necessary to constrain the photospheric luminosity and temperature of L\,1014-IRS
to confirm that this source is substellar.  Preliminary attempts to fit the L\,1014-IRS SED
using the radiative transfer model of Whitney et al. (2004) suggest the temperature is
less than 2600~K, but we defer discussion of SED fitting to a more detailed analysis.

The current data available for L\,1014-IRS are consistent with it having a remarkably low
luminosity and perhaps a substellar mass.  However, given the significant amount of scattering
along the line of sight, the protostellar temperature
is not yet constrained, except that its temperature may be hotter than the previous
estimate of 700~K and cooler than 4000~K.  Without further analysis, or near-infrared
spectroscopic observations, a higher temperature low-mass protostar cannot be ruled
out by our observations.

\section{THE L\,1014 CORE}\label{core}

With the case now much stronger for L\,1014-IRS being embedded in the 
nearby L\,1014 core, we make use of our near-infrared observations 
to derive some physical properties of its parental core to compare with those of
other cores.  An advantage of
deep near-infrared observations toward molecular clouds is in the ability to use them
to construct extinction maps with high spatial resolution (e.g., 20\arcsec\ as in Lada et
al. 2004), which can be used to derive a mass for the core.  The extinction profile
for L1014 can be modeled to estimate its density structure.  This density structure can
be compared with those of other starless and star-forming cores in order to gain insight
into the evolution of collapsing cores as they form protostars.  If future observations
confirm L\,1014-IRS to be substellar, then the density structure of
L\,1014 and other cores harboring embedded brown dwarfs may be
particularly interesting in comparing brown dwarf and protostellar formation.

\subsection{High Resolution Extinction Map}\label{map}

\begin{figure*}[t]
\epsscale{1.0}   
\plotone{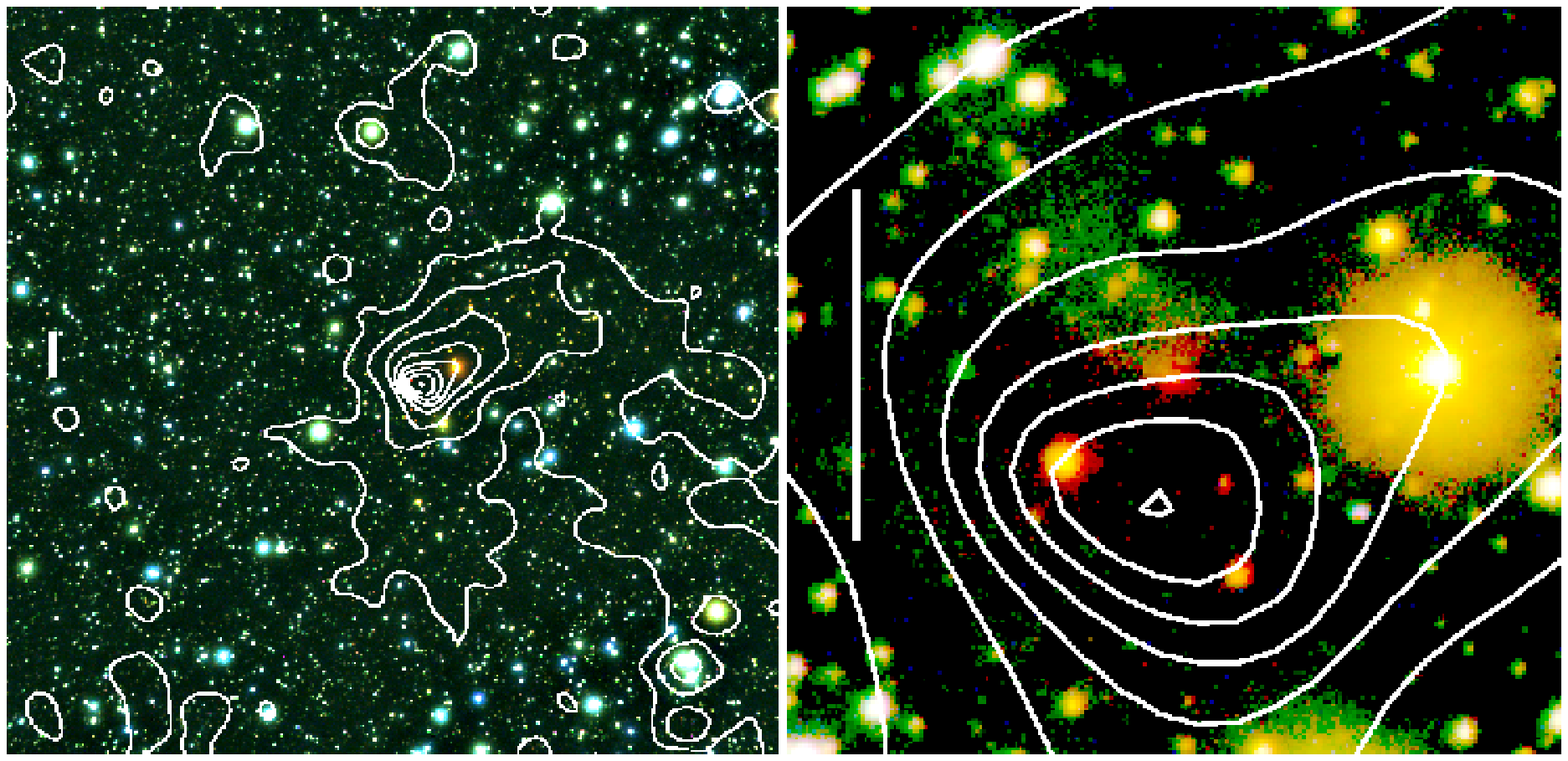}
\caption{JHK$_s$ composite images of the L\,1014 region observed with FLAMINGOS on the KPNO 4-meter telescope
are displayed with beam-averaged extinction contours.  The left panel shows the $\sim$8\arcmin$\times$9\arcmin\ 
(0.48 pc $\times$ 0.52 pc) 
field centered on L1014-IRS, while the right panel shows an enlarged display of only the central part of the core.
The angular scale is represented by the 30\arcsec\ (0.03 pc) white bar along the left side of each image.
The extinction was determined using the {\it{NICE}} method (see \S{\ref{map}}) and convolving the line-of-sight
extinction measurements with a 20\arcsec\ Gaussian beam.  In the left panel, contours are overlaid at A$_V$ = [2, 5, 10, 15, 20, 25, 30, 35] magnitudes, with A$_V$ = 2 magnitudes representing the
2$\sigma$ noise level in these extinctions.
The same contours appear in the right panel, with the exception of the A$_V$ = 2 magnitude contour.
\label{L1014_JHK_ext}}
\end{figure*}

Differential extinction as a function of wavelength gives rise to interstellar
reddening, resulting in observed colors of stars that are redder than
their intrinsic colors.  This difference, known as a color excess, for $H-K_s$
is explicitly written as

\begin{equation}
E(H-K_s) = (H-K_s)_{observed} - (H-K_s)_{intrinsic} .
\label{excess_eqn}
\end{equation}

\noindent
Column densities 
mapped by this method, known as the {\it{NICE}} ({\it{N}}ear-{\it{I}}nfrared {\it{C}}olor {\it{E}}xcess; Lada, Alves, \& Lada 1999) method, 
historically have been expressed in terms of their equivalent visual extinctions, $A_V$. 
Given the observed reddening law of Rieke \& Lebofsky (1985), the visual extinction
is related to E(H-K$_s$) by

\begin{equation}
A_V = 15.9\ E(H-K_s) .
\label{av_eqn}
\end{equation}

The FLAMINGOS field is sufficiently large that
stars just beyond the core boundaries (e.g., regions exhibiting no color
excess or cloud structure) may be used to determine the intrinsic colors
of the background star population toward L\,1014.  The region used to compute the
intrinsic colors is given by (2000.0)

\begin{eqnarray}
21^h24^m28.2^s & <\ \ \alpha & < 21^h24^m37.1^s  \\
49^\circ 59^\prime 13^{\prime\prime} & <\ \ \delta & < 50^\circ 01^\prime 34^{\prime\prime}
\end{eqnarray}

\noindent
Within that region, the photometry of 247 stars with H and K$_s$ photometric
uncertainties less than 0.20 magnitude was used to obtain
an intrinsic color of 

\begin{equation}
(H-K_s)_{intrinsic} = 0.199 \pm\ 0.009 .
\end{equation}

\noindent
The uncertainty quoted is the error in the mean H$-$K$_s$ color.
The dispersion in the H$-$K$_s$ colors of these non-extincted stars
was 0.149~magnitude and represents the photometric
uncertainties in addition to the uncertainty in the intrinsic color of an
individual background star.  Thus, an individual line-of-sight extinction
is uncertain by $\sim$2.4~magnitudes.  Since the photometric
uncertainty depends on the brightness of the star, and the background
star population will tend to be fainter than the non-extincted stars,
the uncertainty in an individual line-of-sight extinction through
the densest regions may be closer to $\sim$3 magnitudes.

With the intrinsic H$-$K$_s$ color, the observed H$-$K$_s$ colors of stars
(with H and K$_s$ photometric uncertainties less than 0.20 magnitude)
were converted to line-of-sight extinctions.  Then, in order to reduce the
errors in extinction determinations, these line-of-sight
extinctions were convolved with a Gaussian beam
with FWHM of 20\arcsec\ to produce the uniformly sampled, high-resolution
extinction map shown in Figure~\ref{L1014_JHK_ext}.

The beam-average extinction peaks at A$_V$ = 35 magnitudes at a position,
offset from L\,1014-IRS, as given in Table~\ref{peakposition}.  Since this position was determined from
a 20\arcsec\ beam, the true position of peak extinction is likely to be within
10\arcsec\ of that position.  The peak line-of-sight extinction is
A$_V \sim$ 44 magnitudes, found for three stars south of L\,1014-IRS.  The 
next greatest line-of-sight extinction is A$_V =$ 29 magnitudes.
Making use of line-of-sight extinctions to refine the peak position, we obtain
a position offset 4\arcsec\ east and 14\arcsec\ south, as indicated in
Table~\ref{peakposition}.  This refined position was derived such that the
extinction-weighted distances from the peak to each of the three stars with
greatest extinctions was equal.

The line-of-sight extinctions near the center of L1014, as probed by the {\it{NICE}} method,
are comparable to those derived in similar studies of other isolated dense cores.
For example, in the deep near-infrared imaging surveys of Kandori et al. (2005) and 
Murphy \& Myers (2003), peak extinctions of A$_V \sim$ 30--50 magnitudes were observed
in almost half of their combined sample of 18 dense cores.
Harvey et al. (2003a) observed line-of-sight extinctions of A$_V \sim$ 30 magnitudes
at about 30\arcsec\ from the center of the starless core L694-2.  For the protostellar
core B335, Harvey et al. (2001) observed extinctions of A$_V \sim$ 50 magnitudes
at about 20\arcsec\ from the center.  

Other determinations of peak column density positions, derived from dust emission maps
of L\,1014, are included in Table~\ref{peakposition}.  These positions are all south of
L\,1014-IRS, but north of the peak position as determined from our dust extinction study.
Since the dust emission is a function of the column density and temperature, it is not
surprising that the dust emission peaks closer to L\,1014-IRS than our extinction
peak.  L\,1014-IRS heats its circumstellar dust and, given the beamsize of the (sub)millimeter observations, will shift the dust emission peak toward L\,1014-IRS and away from
the column density peak.

The 10--15\arcsec\ offset between L\,1014-IRS and the column density peak, as given
by our extinction observations, may be explained in several ways.  In Smoothed Particle
Hydrodynamic (SPH) simulations, Stamatellos et al. (2005) demonstrate that protostars
may be displaced from the core center due to asymmetries in the pattern of accretion.
In one of their simulations, the protostar was given a peculiar velocity of $\sim$0.3~km~s$^{-1}$,
comparable to the projected peculiar velocity of 0.1~km~s$^{-1}$ for L\,1014-IRS, assuming
an age of 10$^5$ years.  According to these models, a consequence of this displacement
is a significantly reduced accretion rate, slowing the growth of the protostar.  We speculate
that, if future observations of L\,1014-IRS demonstrate it to be substellar, one difference
between the formation of protostars and brown dwarfs may be that the accretion of
a brown dwarf may have been turned off prematurely due to higher peculiar velocities
imparted by asymmetric accretion.  In this case, embedded brown dwarfs should be 
observed with greater offsets from their parental core centers compared to 
protostars.  An alternative explanation for the offset between L\,1014-IRS and the extinction
peak may be that the structure of the L\,1014 core includes two distinct condensations
that, in projection, appear close together.  One of these condensations formed
L\,1014-IRS, while the other remains intact.

\begin{table*}[bth]
\caption{Determinations of Peak Column Density}
\label{peakposition}
\begin{center}
\begin{tabular}{lcccc}
\hline
\hline \\[-9pt]
									&										& \twocolheads{Offset from L\,1014-IRS}		& Peak ${\cal{N}}_H$  \\
Observations   							& Beamsize								& \ \ \ $\Delta$ RA		& \ \ \ $\Delta$ Dec.			& [10$^{22}$ cm$^{-2}$]	\\[3pt]
\hline \\[-9pt]
SCUBA 850-$\mu$m Dust Emission Map$^a$	& 16$^{\prime\prime}$						& $+$0$^{\prime\prime}$	& $-$6$^{\prime\prime}$		& 4.3		\\
MAMBO 1.2-mm Dust Emission Map$^{b,c}$	& 20$^{\prime\prime}$						& $-$1$^{\prime\prime}$   & $-$3$^{\prime\prime}$		& 3.5		 \\
BIMA 3-mm Dust Emission Map$^d$		& 22$^{\prime\prime}\times$16$^{\prime\prime}$	& $+$3$^{\prime\prime}$	& $-$7$^{\prime\prime}$		& ...		\\
{\it{NICE}} Dust Extinction Map				& 20$^{\prime\prime}$						& $+$2$^{\prime\prime}$	& $-$10$^{\prime\prime}$		& 6.3		\\
{\it{NICE}} Line-of-Sight Dust Extinctions		& $<$20$^{\prime\prime}$					& $+$4$^{\prime\prime}$	& $-$14$^{\prime\prime}$		& 7.9		\\
\hline
\hline
\end{tabular}
\end{center}
\footnotesize{$^a$ Visser, Richer, \& Chandler (2002); Young et al. (2004)} \\
\footnotesize{$^b$ Young et al. (2004); Kauffmann et al. 2005, in preparation.} \\
\footnotesize{$^c$ Original data with 11\arcsec\ beam were used to derive peak column density, but original data were smoothed to 20\arcsec\ to derive position.} \\
\footnotesize{$^d$ Lai et al. 2005, in preparation.; Position was reliably determined, but peak column density could not be derived reliably since these observations resolved out some of the core emission.} \\
\end{table*}

\subsection{Core Mass}\label{coremass}

The {\it{NICE}} method of mapping dust extinction
provides a relatively reliable determination of the mass of the molecular
cloud, independent of an assumed geometry for the cloud.  The gas-to-dust
ratio widely used is that derived by the study of Bohlin, Savage, \& Drake (1978),
which determined the column density of hydrogen in the diffuse interstellar medium
(ISM) as a function of the optical color excess $E(B-V)$.  Making use of the
standard reddening law (Rieke \& Lebofsky 1985), their gas-to-dust ratio
can be expressed in terms of visual extinction $A_V$ as

\begin{figure}[tb]
\epsscale{0.5}   
\begin{turn}{90}
\plotone{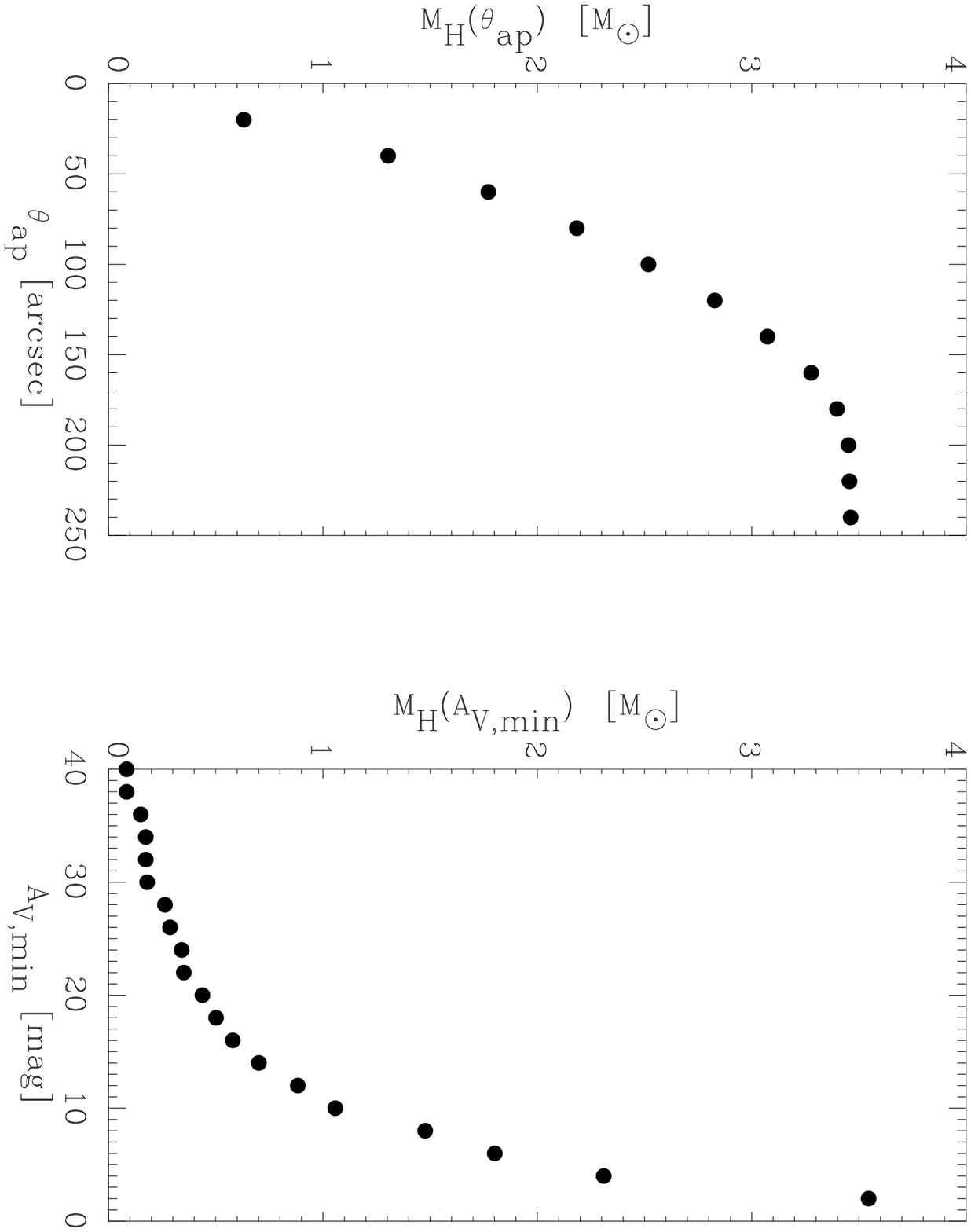}
\end{turn}
\caption{
Cumulative mass profiles for the L\,1014 core, derived using the {\it{NICE}} method (see \S\ref{coremass}).  In each case, statistical error bars, which are typically $\sim$1-2\% of the
mass, are smaller than the symbols.  [Left]  Mass as a function of the radius of circular apertures centered on the position of peak extinction.  [Right]  Mass as a function of the smallest extinction
value considered.  This profile effectively uses the elongated, irregular extinction contours as apertures
rather than circular apertures.  The mass in regions with $A_V \ge$ 2 magnitudes (the
2$\sigma$ noise level in our extinction map) is comparable to the mass within the largest circular
apertures plotted to the left.
\label{massrad}}
\end{figure}

\begin{equation}
{\cal{N}}_H = 1.8 \times 10^{21} A_V\ \ \ {\rm{atoms}}\ {\rm{mag}}^{-1}\ {\rm{cm}}^{-2}
\label{bohlin_eq}
\end{equation}

\noindent
The mass of the molecular
cloud is determined by integrating the mass column density over the
area of the cloud,

\begin{equation}
M_H = m_H\ \int_\Omega {\cal{N}}_H  d\Omega ,
\end{equation}

\noindent
where $m_H$ is the mean molecular weight of gas within the core and $\Omega$ is the projected area of the cloud on the sky.

Using circular apertures of different radii centered on the position of peak extinction, we plot in
Figure~\ref{massrad} the mass of the L1014 core as a function of aperture radius $\theta_{ap}$,
determined by

\begin{equation}
M_H(\theta_{ap}) = 2 \pi\ (1.8 \times 10^{21})\ m_H\ D^2\ \int_0^{\theta_{ap}} (A_V - A_{V,bg} ) \ \theta\  d\theta ,
\label{massint_eq}
\end{equation}

\noindent
where $D$ is the distance and $A_{V,bg}$ is the component of extinction attributed to
the more extended component of the L\,1014 cloud within which the core is embedded (or, alternatively,
$A_{V,bg}$ represents a systematic correction to our derived intrinsic $H-K$ color, as explained in \S\ref{physical_properties}).  
This plot demonstrates that the mass asymptotically approaches 3.46 $\pm$ 0.03~M$_\odot$ with increasing aperture size.  Systematic uncertainties in these masses, dominated by uncertainties in the distance, are not included in our mass estimates.

For comparison, Young et al. (2004) used the 850~\micron\ dust emission map of L\,1014 to derive
a mass of 1.7~M$_\odot$ within 75\arcsec\ (15000 AU) of the center of the core.
Using Equation~\ref{massint_eq}, we derive a mass of 2.1~M$_\odot$ within
the same region.  This difference in mass determination is likely due to uncertainty in the
submillimeter opacity or the reddening law for such dense environments, or both.

Most of the mass of the core is within the more extended regions of low column density.
This fact is most evident when plotting the cumulative mass as a function of minimum
extinction, $A_{V,min}$, as in Figure~\ref{massrad}.  In particular, the mass of the core in regions
with extinction $A_V \ge$ 10 magnitudes is only 1.1~M$_\odot$, while the mass in
regions with extinction $A_V \ge$ 2 magnitudes is more than triple that, at 3.55~$\pm$ 0.02 M$_\odot$.

\subsection{Physical Properties from Extinction Modeling}\label{physical_properties}

For comparison with previous studies, we first model the L\,1014 core as an isothermal
spherical core in hydrostatic equilibrium, known as a Bonnor-Ebert core.  Since L\,1014 is clearly
elongated, having an aspect ratio of $\sim$2, we also consider the case of an isothermal
cylindrical core for comparison.  L\,1014 has a young embedded source and hence at least some portion
of the core is likely to be collapsing and therefore not in hydrostatic equilibrium.  For this reason,
the modeling of L\,1014 in this study is not for the purpose of ascertaining the hydrodynamic stability
of the core.  Instead, we make use of the density profiles derived from such modeling since they,
regardless of stability, represent a core structure consistent with the observations.
Furthermore, models of collapsing cores with Bonnor-Ebert initial states have density profiles very
similar to Bonnor-Ebert profiles for the early stages of collapse (Myers 2005).
Models of Bonnor-Ebert and isothermal cylindrical cores have been
used recently to characterize B335 (Harvey et al. 2001), the prototypical Class~0 protostellar
core.  Starless cores, such as B68 (Alves, Lada, \& Lada 2001), L\,694-2 (Harvey et al. 2003a),
and L\,1544 (Evans et al. 2001) also have been fitted to such models.  By comparing the density
structures of different types of cores, we might be able to construct an evolutionary sequence
of dense cores as they collapse to form low-mass protostars or brown dwarfs.

\begin{figure*}[t]
\epsscale{0.75}   
\plotone{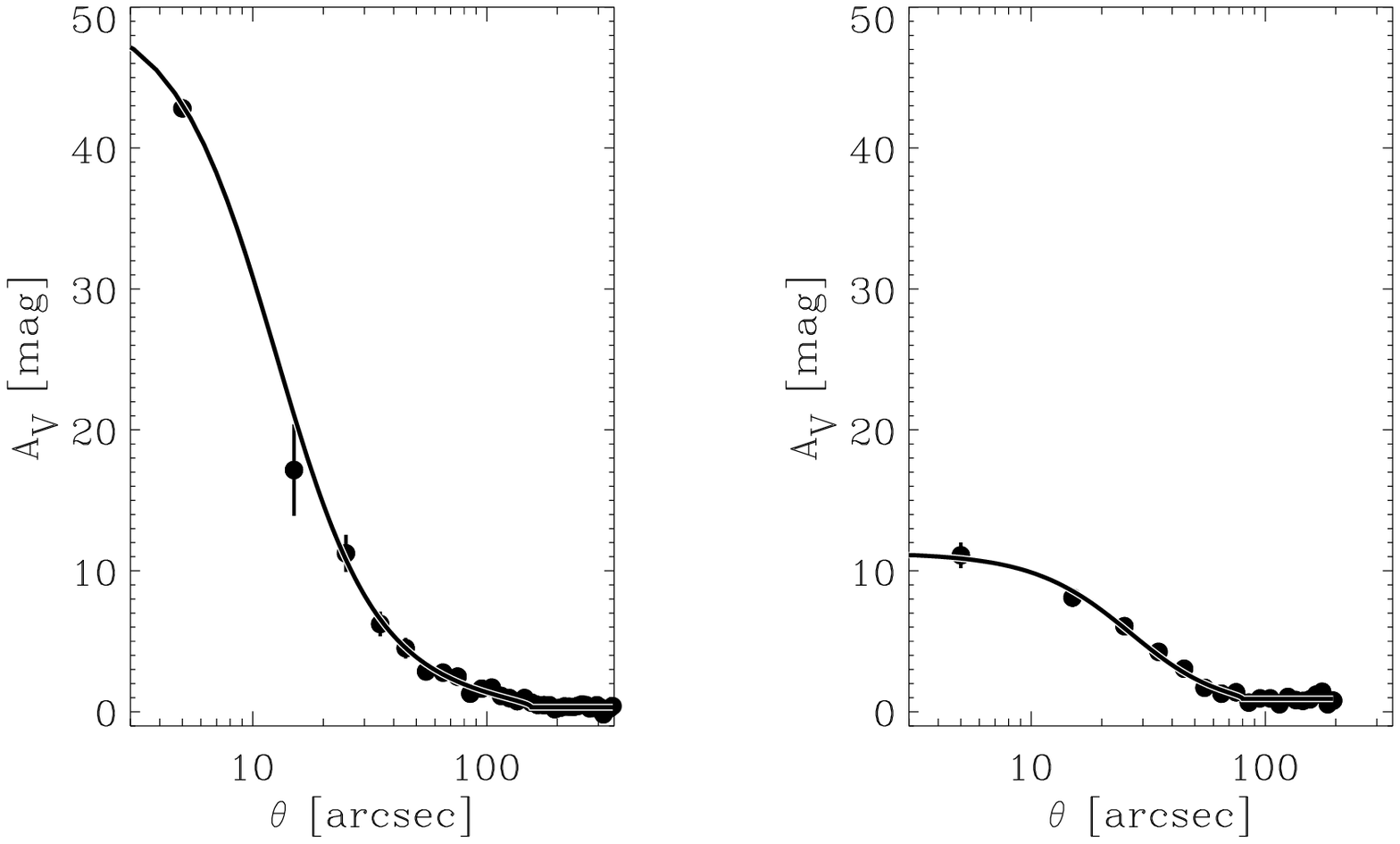}
\caption{Observed extinction profiles for the L\,1014 core with the best isothermal fits overlaid.
[Left] The data points represent the azimuthally averaged, radial profile of extinction, making use of 10\arcsec\ shells.  The best-fit Bonnor-Ebert model, plotted as a solid curve, matches the
observed profile well with $\chi^2_{red} \approx\ 0.7$.  [Right] The cylindrically averaged extinction
profile, making use of 10\arcsec\ slabs, is represented by the data points.  Similar to the Bonnor-Ebert
core, the best-fit isothermal cylindrical filament, plotted as a solid curve, is consistent with the observations.
\label{profiles}}
\end{figure*}

Adopting the peak position determined from the line-of-sight extinctions
given in Table~\ref{peakposition}, we construct a radial profile of extinction
for L\,1014 in Figure~\ref{profiles} by averaging the line-of-sight extinctions of stars within 
10\arcsec-wide shells centered on the peak position.  With this shell size,
the central shell contains the three stars with the greatest (and similar) line-of-sight
extinctions.  The second smallest shell (outer radius of 20\arcsec) contains five
stars.  For larger shells, the number of stars is nearly proportional to the area, with
30 stars found within the third smallest shell (outer radius of 30\arcsec).
The error bar on each data point
represents the error in the mean of the line-of-sight extinctions within each shell.
In general, these error bars are expected to include the photometric errors as well as 
errors in the assumed intrinsic color.  Beyond $\sim$200\arcsec, the mean extinction
derived from the azimuthally averaged profile has not quite reached $A_V = 0$ magnitude.
Rather the profile appears to hover near $A_V \approx\ 0.3$ magnitude, which could be
attributed to a more extended cloud component within which the core is embedded.
However, given that this ``background'' level of extinction
can be explained instead by increasing the intrinsic $H-K_s$ color, determined in
\S\ref{map}, by only 2$\sigma$, this apparent extended cloud component may simply
represent a systematic error caused by a slight offset in the intrinsic color.  The results
of our study do not depend on the particular explanation since we are mainly concerned
with that portion of the extinction due to the core itself.

Details of Bonnor-Ebert modeling of radial extinction profiles can be found in
Harvey et al. (2001).
In standard fitting of truncated Bonnor-Ebert models to extinction profiles, there are three ``observational'' parameters that are directly constrained by the profile:  $\xi_{max}$, $\theta_{max}$, and
$A_{V,0}$.  The shape of a normalized Bonnor-Ebert profile is characterized by $\xi_{max}$,
with larger values denoting more centrally condensed cores.  The angular radius of the core, $\theta_{max}$, and peak extinction through its center, $A_{V,0}$, enable comparison of these
normalized profiles with observed profiles.  In addition, we consider that the extinction profile of
the core is superimposed onto a constant level of extinction, $A_{V,bg}$, from the more extended
molecular cloud within which the core is embedded.   Finally, in Bonnor-Ebert modeling, 
there is often a weak ``degeneracy'' between $\xi_{max}$ and $\theta_{max}$, especially if
the density profile is very centrally condensed and there are no extinction observations
sufficiently close to the core center.  In such cases, greater values
of $\xi_{max}$ may provide fits nearly as good as smaller $\xi_{max}$ values,
provided $\theta_{max}$ is correspondingly greater.  Yet, greater values of
$\xi_{max}$ and $\theta_{max}$ imply smaller external pressures on the
boundary of the core.  While fits with large values of $\xi_{max}$ and $\theta_{max}$
may provide good fits to the extinction profile, the implied external pressure is
unrealistically too small.  The pressure in the local interstellar medium is
$P_{ext} / k \sim~2~\times~10^4$  K\ cm$^{-3}$ (Boulares \& Cox 1990),
while external pressures confining molecular clouds and cores are typically a factor
of a few, perhaps as large as 10, greater than this value (e.g., McKee \& Holliman 1999).
Similar to considerations made
by Zucconi, Walmsley, \& Galli (2001), we constrain the external pressure in our
Bonnor-Ebert modeling to be greater than the minimum value, $P_{ext} / k > 2~\times~10^4$  K\ cm$^{-3}$.

Normalized, theoretical extinction
profiles of Bonnor-Ebert cores were generated for a grid of ($\theta_{max}$, $\xi_{max}$) values for
$\theta_{max} = [50^{\prime\prime},350^{\prime\prime}]$,
incremented by 5$^{\prime\prime}$, and $\xi_{max} = [1,50]$,
incremented by 0.1.  Each of these normalized theoretical profiles were
then fitted to the observed profile by optimizing $A_{V,0}$ and $A_{V,bg}$,
while ensuring the external pressure constraint is satisfied.  The quality of
these fits were judged by the standard reduced chi-squared values,
$\chi^2_{red}$.  In order for a model to satisfactorily agree with observations,
typically $\chi^2_{red} \ltsim\ 1$ is required.

The best Bonnor-Ebert fit to the L\,1014 radial extinction profile
was consistent with observations, yielding $\chi^2_{red} = 0.7$.
The observational parameter values corresponding to this
fit are listed in Table~\ref{isothermal_sphere_params} and
the corresponding extinction profile is overlaid onto the observed
profile in Figure~\ref{profiles}.  We also list in Table~\ref{isothermal_sphere_params}
the 99\% confidence ranges, representing the set
of parameter values for which

\begin{equation}
\chi^2_{red} \le 1 + 3/n_{free} ,
\label{chi2red_99perc}
\end{equation}

\noindent
where $n_{free}$ is the number of degrees of freedom.

\begin{table*}[tb]
\caption{Fitted Parameters}
\label{isothermal_sphere_params}
\begin{center}
\begin{tabular}{lcccc}
\hline
			& \twocolheads{Bonnor-Ebert}						& \twocolheads{Isothermal Cylinder}	\\
Parameter   	& Best-Fit Value	& 99\% Confidence Range		& Best-Fit Value	& 99\% Confidence Range	\\[3pt]
\hline \\[-9pt]
$\xi_{max}$						& 35.8	& 29.2 -- 36.9		& 6.9		& 3.2 -- 9.2	\\
$\theta_{max} [arcsec]$				& 155	& 135 -- 165		& 80		& 55 -- 100		\\
A$_{V,0} [mag]$					& 50.2	& 47.7 -- 51.8		& 10.4	& 8.2 -- 12.8	\\
A$_{V,bg} [mag]$					& 0.3		& 0.2 -- 0.4		& 0.9		& 0.8 -- 1.0	\\
\hline
\hline
\end{tabular}
\end{center}
\end{table*}

From Table~\ref{isothermal_sphere_params}, we see that a range of 
$\xi_{max}$ values, given by $29.2 \ltsim\ \xi_{max} \ltsim\ 36.9$, were
found to satisfactorily agree with our observations of L\,1014.  Previously,
the starless core B68 was found to be characterized by a density
profile given by $\xi_{max} = 6.9 \pm\ 0.2$ (Alves, Lada, \& Lada 2001);
the density profile of the well-studied protostellar core B335 was found to
be more centrally condensed than that of B68 and characterized by
$\xi_{max} = 12.5 \pm\ 2.6$ (Harvey et al. 2001).  From our analysis of
the dust extinction profile presented here, L\,1014 is more centrally
condensed than both of these prototypical starless and star-forming cores.

In fact, L\,1014 is more centrally condensed than would be inferred from
previous dust emission (Visser, Richer, \& Chandler 2002; Kauffmann et al. 2005, in preparation.;
Lai et al. 2005, in preparation) and molecular line observations (Crapsi et al. 2005a, 2005b).
For example, the integrated intensity of N$_2$H$^+$(1-0) at the peak
of L\,1014 is almost a third that of the L\,694-2 core.
Yet, L\,1014 is ${\it{more}}$ condensed than L\,694-2, characterized by
$\xi_{max} = 25 \pm\ 3$ from deep near-infrared observations of extinction
(Harvey et al. 2003a).  Referring to the azimuthally averaged radial extinction
profile for L\,1014 in Figure~\ref{profiles}, we see that the column density of L\,1014
falls to half its peak value within $\sim$10\arcsec, whereas this characteristic size
of L\,694-2 is 20--30\arcsec.  With beamsizes of $\sim$20\arcsec, single-dish
millimeter observations cannot sample with sufficient angular resolution
such a small, highly condensed core as L\,1014.

Herein, we find the advantage of near-infrared observations
and the {\it{NICE}} method in that individual lines of sight arbitrarily close to the peak
column density may be sampled by deeper and deeper observations.  For
purposes of studying the inner structure of such small cores like L\,1014, it 
is necessary to achieve angular resolutions provided by deep near-infrared
observations or interferometric observations (e.g., Harvey et al. 2003b, 2003c;
Bourke et al. 2005) of dust emission or molecular lines.  Indeed, it is possible
that isolated regions of low-mass star formation, or brown dwarf formation, may
not have been detected by previous single-dish millimeter observations.

On the other hand, L\,1544 and L\,1689B are two examples of isolated starless cores,
for which previous single-dish observations of dust emission (Evans et al. 2001)
indicate that they are very centrally condensed ($\xi_{max} \approx\ 20-50$)
and should be forming stars.  L\,694-2 is another example.
However, previous observations of L\,694-2, L\,1544, and L\,1689B were not
sensitive enough to detect very low-mass protostars or embedded substellar
candidates such as L\,1014-IRS.
If the density structure is a significant factor in forming low mass protostars (e.g.,
Kandori et al. 2005; Muench et al. 2005, in preparation) and substellar sources, then we might
expect that more sensitive near-infrared or mid-infrared observations of these
cores might reveal embedded sources, which may be substellar.

Given the elongated shape of L\,1014, for comparison with the radial extinction
profile, we generate the extinction profile as a function of angular distance from
the major axis, shown in Figure~\ref{profiles}.
Comparing this cylindrical profile with the radial profile,
one striking difference is that the peak extinction in the radial profile is 
significantly greater than that in the cylindrical profile.  This difference is due
to the variation of the line-of-sight extinctions along the major axis, effectively
diluting the peak extinction in the profile.  Another difference between the radial
and cylindrical profiles is that L\,1014 appears to extend to a radius
$\theta_{max}\sim$155\arcsec\ when assuming a spherical symmetry, whereas
it extends about half that distance in the cylindrical profile.  The size of the core as given
by the spherical profile will closely correspond to its major axis, while the extent of 
the cylindrical profile should closely correspond to the size of its minor axis.
Hence, these profiles are consistent with L\,1014
having an aspect ratio of $\sim$2:1, as apparent from the extinction contours
in Figure~\ref{L1014_JHK_ext}.

Using a method similar to that for Bonnor-Ebert cores (e.g., Harvey et al. 2003a),
we find that the cylindrical profile can be fit to models of truncated isothermal cylindrical
filaments (Ostriker 1964).  The best-fit values and the 99\% confidence ranges for the observational
parameters are listed in Table~\ref{isothermal_sphere_params}.  With the
best-fit cylindrical filament model yielding $\chi^2_{red} = 0.6$, these models
fit our observations equally well compared to Bonnor-Ebert cores.

The density profiles given by some isothermal cylindrical core models and very centrally condensed
Bonnor-Ebert core models are consistent with the L\,1014 extinction profiles.  These models 
constrain certain physical properties of L\,1014, namely $T$ and $n_0$, as well as the
distance.  Since at least
some of the L\,1014 core is not in hydrostatic equilibrium, the temperature that is inferred by
such modeling may not be directly applicable to L\,1014.  For this reason, we refer to the
temperature inferred by the modeling as an {\it{effective}} temperature.
With that note of caution, we briefly investigate the implications of the modeling on
$T_{\rm{eff}}$, $n_0$, and $D$.
The extent to which these inferred quantities are valid for
L\,1014 depends on the process by which star formation occurs in dense cores
and the portion of the core affected.

The physical parameters are related to the fitted observational parameters
by the following two constraints:

\begin{equation}
D_{pc} \sqrt{\frac{n_c}{T_{\rm{eff}}}} = 2.2 \times\ 10^5 \left( \frac{\xi_{max}}{\theta_{max}} \right)  {\rm{pc\ cm^{-3/2}\ K^{-1/2}}} ,
\label{BEconst1}
\end{equation}

\begin{equation}
n_c T_{\rm{eff}} = 3.2 \times\ 10^5 \left[ \frac{A_{V,max}}{\xi_{max} \Lambda(\xi_{max})} \right]^2   {\rm{mag^{-2}\ cm^{-3}\ K}} ,
\label{BEconst2}
\end{equation}

\noindent
where $\Lambda(\xi_{max})$ is the dimensionless column density through the center of the
core,

\begin{equation}
\Lambda(\xi_{max}) = \frac{2}{\sin{\phi}} \int_0^{\xi_{max}} \frac{n(\xi, \xi_{max})}{n_c} d\xi ,
\label{lambdaEQ}
\end{equation}

\noindent
computed by numerical integration.  Cylindrical filaments involve one more parameter than
spherical cores --- the inclination of the core relative to the line of sight, designated as $\phi$
in Equation~\ref{lambdaEQ}.  However, the extinction profile does not constrain this
inclination.  In the case of a Bonnor-Ebert core, $\phi\ = 90^\circ$ in Equation~\ref{lambdaEQ}.
Assuming a distance of 200~pc for L\,1014,
the 99\% confidence range for the Bonnor-Ebert core results in an effective temperature of
18--22~K and central density of 1.0--1.4 $\times$ 10$^6$ cm$^{-3}$, as summarized
in Table~\ref{physpropsD200}.  In this table, we also include the range of
core radii $R$, derived from the distance and angular radii inferred from this model.
For comparison, the quantities computed from the
isothermal cylindrical core model are also listed in Tables~\ref{physpropsD200},
with $R$ in this case referring to the maximum extent of the L\,1014 filament from 
the major axis.

Comparing the physical properties of the L\,1014 core as inferred from the Bonnor-Ebert
modeling with those inferred from the isothermal cylindrical core, 
the spherical core has a greater central density than the cylindrical core by an
order of magnitude.  This difference likely can be explained by the Bonnor-Ebert
directly modeling the peak central density, probed by the peak central extinction, 
whereas the isothermal cylindrical core model considered here is only considering 
the average density along the major axis.  Clearly, the extinction varies along 
this major axis and a more careful treatment of elongated cores should account
for this feature.  Finally, if the distance to L\,1014
is assumed, both the Bonnor-Ebert and cylindrical core models predict similar
effective temperatures for L\,1014.

\begin{table}[tb]
\caption{Derived Physical Properties, with D $=$ 200~pc}
\label{physpropsD200}
\begin{center}
\begin{tabular}{lcc}
\hline
Property   					& Bonnor-Ebert		& Isothermal Cylinder  \\[3pt]
\hline \\[-9pt]
T$_{\rm{eff}}$ [K]			& 18$-$22				& 14$-$20		\\
n$_0$ [10$^6$ cm$^{-3}$]	& 1.0$-$1.4				& (0.1$-$0.2)~$\sin{\phi}$	\\
R [pc]					& 0.13$-$0.16				& 0.05$-$0.10		\\
\hline
\hline
\end{tabular}
\end{center}
\end{table}

\section{CONCLUSION}\label{conclusion}

Deep near-infrared observations presented here have detected the K$_s$-band counterpart
to L\,1014-IRS, the mid-infrared source with protostellar colors detected by the Spitzer Space
Telescope.  At the time of discovery, the harboring core of this protostellar
source was ambiguous --- either L\,1014 at a distance of 200~pc, or a more distant
core associated with the Perseus arm at 2.6~kpc.  The spatial extent of near-infrared
nebulosity reported here, which traces the evacuated bipolar cavity
typically associated with young stellar objects and protostars, strongly suggests
that L\,1014-IRS is indeed associated with L\,1014.  In this case, L\,1014-IRS is the lowest
luminosity source yet detected within a molecular cloud core; comparison with
evolutionary tracks indicates that this source may be substellar.  Given the 
large contamination of scattered light at K$_s$ along the line of sight to L\,1014-IRS,
and the morphology of the bipolar nebula, L\,1014-IRS may be obscured by its 
circumstellar disk.  For this reason, without near-infrared spectroscopic observations
or radiative transfer models extending to lower temperatures,
we cannot rule out the possibility that L\,1014-IRS is instead a deeply embedded,
low-mass protostar.

Despite the lack of any outflow detection in previous, extensive searches,
the near-infrared nebula would seem to be indirect evidence
for an outflow oriented approximately north-south.  Recent interferometric submillimeter
observations from the Submillimeter Array (SMA) have detected the weak
protostellar CO outflow aligned with the near-infrared nebula, confirming that L\,1014-IRS is
an outflow source embedded in the L\,1014 core (Bourke et al. 2005).
This direct detection of an outflow and determination of its characteristics may yield
important information about the earliest stages of star formation or, if L\,1014-IRS is
confirmed to be substellar, elucidate differences between the formation of protostars
and formation of brown dwarfs.

From near-infrared colors of background stars, we derive extinctions along many lines
of sight through the L\,1014 core and construct a high angular resolution extinction
map.  We find that L\,1014 has a peak extinction of $A_V~\sim~50$ magnitudes and
a mass of 3.6~M$_\odot$.  Based on the radial profile of extinction, L\,1014 appears to be
one of the most centrally condensed cores known.  Finally, the extinction map suggests that
L\,1014-IRS may have been displaced by $\sim$10--15\arcsec\ (2000--3000~AU)
from the core center, perhaps due to peculiar velocities imparted by asymmetric
accretion.  If future observations of L\,1014-IRS confirm it to be substellar,
the significant decrease in accretion rate resulting from its displacement
may be a natural explanation for its substellar mass.  With the successful launch
of the Spitzer Space Telescope, which is sensitive to detecting young embedded
brown dwarfs, this hypothesis may be tested by determining offsets between
the embedded brown dwarfs and the core centers and comparing these offsets
with those of low-mass embedded protostars.

\acknowledgments

Support for this work, part of the Spitzer Legacy Science Program, was provided by
NASA through contract 1224608 issued by the Jet Propulsion Laboratory, California
Institute of Technology, under NASA contract 1407.  C.J. Lada and T.L. Huard acknowledge
the support of NASA Origins grants NAG-9520 and NAG-13041 in developing the code for
Bonnor-Ebert and isothermal cylindrical core modeling.  L.J. Crews acknowledges
the support of a faculty research grant from the University of Tennessee at Martin
for his participation in the KPNO observing run.  Finally, T.L. Huard thanks 
R. Gutermuth for kindly providing his {\it{PhotVis}} IDL photometry package,
S. Mohanty for helpful discussions of the evolutionary tracks of protostars
and substellar objects, and K. Stapelfeldt for reading through a draft of this
paper and providing useful suggestions for improvement.

\end{document}